# Robust corner and tangent point detection for strokes with deep learning approach

Long Zeng*, Zhi-kai Dong, Yi-fan Xu

Graduate school at Shenzhen, Tsinghua University

**Abstract**: A robust corner and tangent point detection (CTPD) tool is critical for sketch-based engineering modeling. This paper proposes a robust CTPD approach for hand-drawn strokes with deep learning approach, denoted as CTPD-DL. Its robustness for users, stroke shapes and biased datasets is improved due to multi-scaled point contexts and a vote scheme. Firstly, all stroke points are classified into segments by two deep learning networks, based on scaled point contexts which mimic human's perception. Then, a vote scheme is adopted to analyze the merge conditions and operations for adjacent segments. If most points agree with a stroke's type, this type is accepted. Finally, new corners and tangent points are inserted at transition points. The algorithm's performance is experimented with 1500 strokes of 20 shapes. Results show that our algorithm can achieve 95.3% for all-or-nothing accuracy and 88.6% accuracy for biased datasets, compared to 84.6% and 71% of the state-of-the-art CTPD technique, which is heuristic and empirical-based.

## 1. Introduction

This work originates from our sketch-based engineering modeling (SBEM) system [1], to convert a conceptual sketch into a detailed part directly. A user friendly SBEM should impose as fewer constraints to the sketching process as possible [2]. Such system usually first starts from hand-drawn strokes, which may contain more than one primitives (e.g. lines and curves). Then, these primitives are replaced by mathematically defined precise shapes, for downstream modeling. Strokes for engineering parts usually contain lots of corners and tangent points, which are used to define shape parameters. The robustness of a CTPD tool affects the computation of shape parameters. Therefore, a robust CTPD approach is an important tool for a SBEM system [3-5].

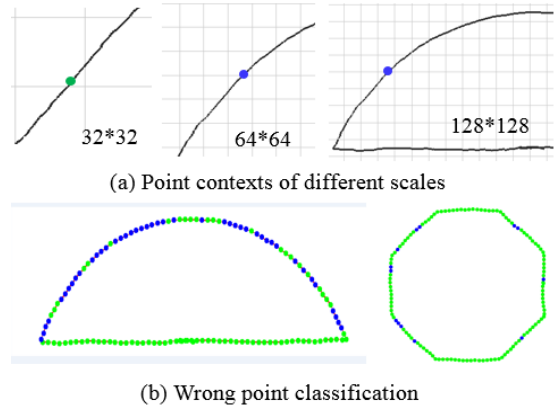

Figure 1. Scale-dependent and classification problems (green: line-points; blue: curve-points)

The robustness of current CTPD techniques is not satisfied due to three reasons. First, the widely used geometric properties, such as straw or curvature (adopted by shortStraw [6] and

iStraw[7]), are local dependent and thus easily affected by scales of *a point context*, i.e. point neighborhood in pixels. The point shown in Figure 1a may be classified as a line-point if its straw value is computed in 32*32 context. In fact, it is a curve-point when judged from larger contexts, such as 64*64 or 128*128. Thus, we argue that if a point is viewed from multi-scaled contexts, the classification should be more accurate [8]. The second is their tolerance to errors, e.g. point classification error. This is inevitable since a hand-drawn stroke is usually noisy and inaccurate, as shown in Figure 1b. Thus, a robust detection algorithm should be tolerant to such kind of classification errors. Third, a CTPD tool should have a stable generalization ability, which means it is unbiased to strokes from different persons, stroke shapes, and datasets. Datasets come from different sources may have quite different noise distribution. Current techniques, e.g. Pu's method [9] and TCVD-R [10], tested their accuracy in their own datasets, which are collected with different user instruction. These datasets with different noise distribution affect the geometric properties' computation and final accuracy. Thus, for practical usage, a CTPD tool should have a good generalization ability.

This paper proposes a robust CTPD approach for hand-drawn strokes with deep learning approach, denoted as CTPD-DL. Firstly, a benchmark dataset with 1500 strokes containing corner and tangent points are collected and labelled. It is used to train two residual neural networks and to evaluate our algorithm. Secondly, multi-scaled point contexts is adopted to classify each point. The point type is determined by its parent primitive. Stroke points are classified into line-points or curve-points by the Point-ResNet, which is an image-oriented residual neural network. Similarly, corners are recognized by Corner-ResNet. Both residual neural networks are trained with multi-scaled point contexts. In such way, a stroke is decomposed into short- and long-segments. Thirdly, a vote scheme is adopted to merge segments into primitives. That is, if most of points agree with a stroke's type, the consensus will be accepted. The adjacent patterns, conditions, and merge operation of short- and long-segments are analyzed and used to guide the segment merge process. Finally, the transition points among primitives are labelled as corners or tangent points according to conditions.

The algorithm's performance and accuracy are experimented with 1500 strokes of 20 shapes. The neural network are trained and optimized with 460800 context images. Results show that our algorithm can achieve 95.3% all-or-nothing accuracy and 88.6% for datasets with different distribution, compared to 84.6% and 71% of the state-of-the-art TCVD-R technique. Thus, our approach is more robust.

To our best knowledge, CTPD-DL is the first try to apply deep learning approach to CTPD. Its robustness is verified by our experiments. The paper is organized as follows. The related work is reviewed in section 2. The algorithm overview and the data set collection method are given in section 3. The train data preparation and classification are introduced in section 4, while the corner and tangent point detection method is detailed in section 5. Section 6 gives experimental results and discussion. Conclusion and future work are summarized in section 7.

## 2. Related work

Corner and tangent point detection is a research topic of stroke segmentation. Similar name are dominant point detection [12] or corner finder. That is, a stroke is divided into primitives at feature points, i.e. corners or tangent points [13]. In this paper, our objective is to segment strokes for SBEM, where corners and tangent points are widely exist. Thus, a feature point detection tool is an important tool for SBEM.

The techniques for feature point detection is developing in two aspects: contents and approaches. The contents is extended from corners of polylines, to corners of polylines and curves, to corners and tangents of an arbitrary stroke. The approaches are developed from heuristic methods to machine learning approaches. In this section, the related detection techniques are reviewed from both aspects.

For corners of polylines, Wolin et al. proposed an easy-to-implement and effective corner detection method, ShortStraw [6]. It resamples the input stroke with an equal distance and computes the "straw value" for each point, which actually is a simplification of local curvature. Then, points with straw value below an empirically determined threshold are considered as candidate corners. The segments between two consecutive candidate corners are checked by a line test. Thus, the quality of ShortStraw heavily depends on the empirical threshold. It usually loses corners drawn in a smooth way since curvature data is noisy and not a reliable data source. Sezgin et al. [14] adopted both curvature and pen speed data for segmentation. That was, they thought corners are points of minimum speed and maximum curvature. Speed data are helpful in most times. However, reference demonstrated it inevitably introduce more noise and counteract the positive effect [15]. In addition, the accuracy of ShortStraw and Sezgin's methods drop quickly when it is applied to curves.

For corners among polylines and curves, Xiong and La Viola extended ShortStraw to arc and curves, denoted as iStraw [7]. It discriminates a corner and curve-point by an angle ratio. These angles are formed by two rays starting from this vertex and end at different sample points on each side. This angle ratio will be a near constant value for a real corner and change for a curve-point. iStraw is still based on curvature, time, and angle information to detect real corners. Alternately, Yu an Cai [16] changed our understanding about corners. That is, corners are dividing points of primitives. Thus, it first fits a stroke with primitives. If the fit is poor, the stroke is segmented at the point with highest curvature. This process is recursively repeated until an error threshold is achieved. Similar iterative processes were adopted by Wolin et al. [17] and SpeedSeg [18].

The contents has been extended to include tangent points. The first detection method for corners and tangent points was proposed by Pu and Gur [9]. First, the minimal set of segmentation points was obtained by a greedy algorithm based on a radial basis function. Then, an angle-based rule was applied to reconstruct primitives, to obtain corners and tangent points. This method has a complex mathematical background, not so easy or intuitive to implement. In 2012, Albert et al. [19] presented a tangent and corner vertices detection method (denoted as TCVD). It also found all corners based on local minimum of radius value, which is an inverse of curvature. Then, the points between two consecutive corners are approximated by cubic parametric curves. This process is also repeated until the maximal fitting error is below a given threshold. Finally, the tangent points are inserted between a line and a curve or a pair of curves of opposite curvature sign. TCVD has a critical threshold since it is used to select points with local minimal radius are constant and scale-dependent. This problem was alleviated by their following work, TCVD-R [10]. They proposed a new mathematical condition for corner detection. That is, a quadratic function to discriminate corners from non-corners were found from a set of exemplar points, i.e. radiuses and second derivative of curvatures of corners and non-corners. However, their method is still sensitive to outlier points.

All of above detection approaches are heuristic methods and their parameters are set empirically. To improve the robustness, some researchers have tried several knowledge-based approach , e.g. agents [20], fuzzy knowledge[21], and neural network [22]. Recently, there is a trend to adopt machine learning approach for feature point detection. For example, Rosten and Drummond proposed a machine learning based high-speed corner detection algorithm for images [23]. Herold and Stahovich presented a machine learning approach to segment hand-drawn strokes into lines and arcs, denoted as ClassySeg [11]. This technique begins by identifying a set of candidate segment windows, containing a local curvature maximum and its neighboring points. Then, geometric features for each point in each candidate segment window are computed. These features are used to train a statistical classifier based on decision tree. Finally, this classifier is used to identify real segment points from those candidate segment windows. The machine-learning framework introduced by this paper has a good extensibility. Its experiments demonstrated the success of the application of machine learning approach for stroke segmentation. Thus, ClassSeg represents a movement toward machine learning approach, instead of heuristic-based approaches.

In this paper, we extended the ClassySeg from both content and approach. The content is extended to include corners and tangent points. Corner detection has been discussed a lot, but tangent point detection is less studied. To our knowledge, only TCVD and Pu's methods have a special discussion about tangent point detection. When sketch is used to support engineering modeling, tangent point detection is an absolutely necessary tool. The detection approach is from decision tree to deep learning neural network approach, which is successfully used in image recognition area [24]. In addition, the multi-scaled context image are adopted. The multi-scale properties of points, e.g. curvatures, have been explored and successfully applied to corner detection in previous studies [25].

### 3. Feature point detection framework

In this section, the algorithm framework of CTPD is presented. In addition, the collection of hand-drawn strokes are detailed.

## 3.1. Detection algorithm framework

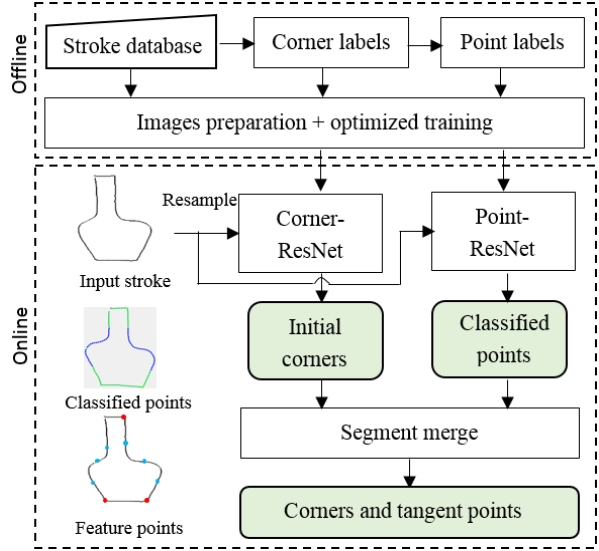

Figure 2. Flowchart of the feature point detection framework

The framework of CTPD-DL algorithm for hand-drawn strokes is given in Figure 2. It consists of two phases: offline and online.

In the offline phase, the objective is to train a residual deep neural network, i.e. ResNet [26]. The input is a stroke database, where corners and points' type are labelled for each stroke. Then, for each point, multi-scaled context images are sampled. The context images of candidate corners and non-corners are separated and divided into a train set and a validate set. When a ResNet is trained with context images of candidate corners, the output is the Corner-ResNet, used to recognize real corners. When a ResNet is trained with context images of non-corners, the output is the Point-ResNet, used to classify points.

In the online phase, the purpose is to detect all corners and tangent points of a hand-drawn stroke. An input stroke is resampled and a set of candidate corners are obtained first. Then, the candidate corners are sent to Corner-ResNet to identify initial corners. Meanwhile, all resampled points are given to Point-ResNet and classified into line-points or curve-points. Finally, the initial corners and classified points are sent to the segment merge algorithm to detect all corners and tangent points.

In the left of Figure 2, the first subfigure is the original input stroke, while the second subfigure is the stroke with classified points. The third subfigure is the stroke with detected corners and tangent points.

## 3.2. Data set

To support the CTPD-DL approach, a large database including 1500 strokes of 20 template shapes with corners and tangent points is created. The 20 template shapes are published by TCVD-R, as shown in Figure 3. The dataset consists of two subsets. One is downloaded from TCVD-R, which includes 1000 strokes. The other 500 strokes are drawn by our research team.

The two data subsets are biased since both have different requirements for participants. We invite 5 participants to redraw the 20 shapes with 5 times for each. After a simple introduction to the drawing environment, we let the participants to look at the shape before to draw it. We do not require each stroke to have the same number of corners and tangent point with the given shape. For example, the strokes in Figure 4a and Figure 4b correspond to the 4[th] and 10[th] template shapes in Figure 3, respectively. The locations 1, 2, and 3 are curves in template shapes, where 6 tangent points should be added. However, in human's perception, i.e. from a wide-angle view, it should be more proper if they are treated as corners.

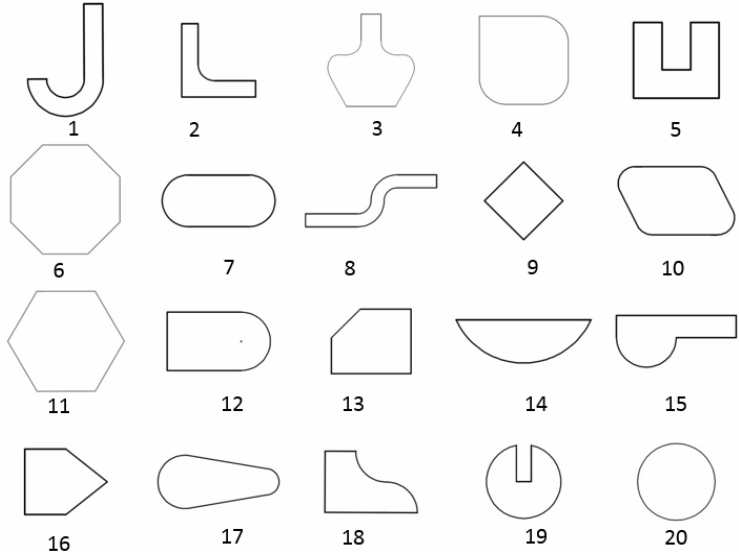

Figure 3. Stroke shapes used in our experiments

Therefore, to be more practical, we further ask users to label the positions of real corners and tangent points for the 1500 strokes, based on human perception [26]. These labels are also used to evaluate the accuracy of our algorithm. This is different with TCVD-R, which compared the detected corners and tangent points with corresponding points of template shapes. Thus, the correspondences between users' strokes and template shapes is not important in our work. However, in TCVD-R, only strokes with correct correspondence are selected. Therefore, this two data subsets have different bias.

Biased dataset also often occur in practical, such as sketch-based image modeling. An image is imported as a background picture and users draw their strokes along the image's contour. Such kind of strokes has quite different noise distribution (such as Gaussian distribution).

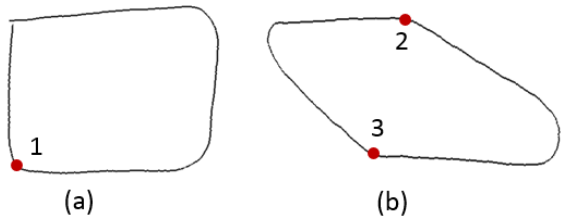

Figure 4. Accepted strokes in our experiment

## 4. Classification with deep learning approach

The image oriented residual neural network—ResNet is trained with different point context images to classify corners and points. ResNet is a supervised learning technique. The training data preparation and point classification are detailed.

### 4.1. Train data preparation

A *stroke* is defined as a sequence of points generated by a pen or a finger, starting from its contact to a touch screen and end by its leave. To simplify the following label and detection process, each stroke is resampled with fixed number of points. A resampled stroke is $stk = \{p_i | i = 1, \cdots, n\}$, where $n$ is equal to 128 in our experiments. Each point is represented by $p_i(x_i, y_i, l_i)$, where $x_i$ and $y_i$ are coordinates; $l_i$ is a label value and has four discrete states: line-point, curve-point, corner, and tangent point. A *line-point* is a point of straight lines. A *curve-point* is a point of curves, e.g. arcs or freeform curves. A *corner* is a point with high curvature, which is a transition point where two primitives meet sharply. A *tangent point* is a transition point where a line and a curve meet smoothly. The discrete value is 0-3 respectively. A stroke may contain more than one lines and curves which are separated by corners or tangent points.

The train data preparation includes stroke label and multi-scaled context image generation.

**Stroke label**

The train purpose is to teach ResNet under what situations a point is a line-point, a curve-point or a corner. Thus, three point types need to be labeled.

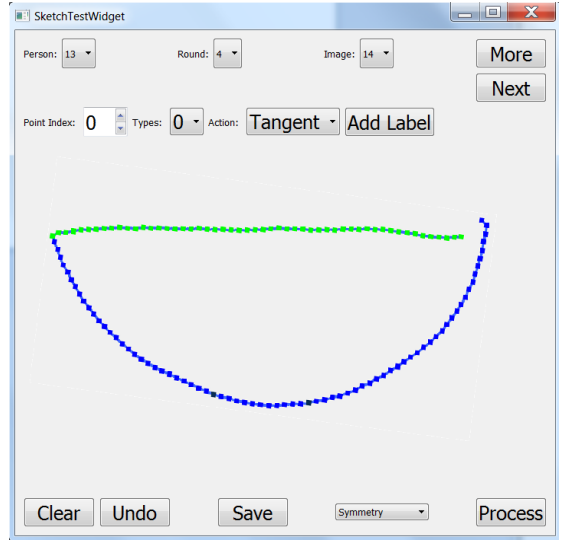

Figure 5. A stroke is labelled into line-points (green), curve-points (blue), and corners (red)

To simplify the label process, a label program is implemented, as shown in Figure 5. First, users open a stroke file. The file name of a stroke is named in the form "person-round-shape.stk". For example, "13-4-14.stk" is a stroke of the 14[th] shape drawn by the 13[th] person in the 4[th] round. Users can select a stroke by configuring the three integers via list boxes (i.e. "Person", "Round", and "Shape") in the label program (Figure 5).

Second, corners are labelled. The number of true corners only occupies a small part of the total points of a stroke. This will cause problem during train process. To balance the number of true positives (i.e. corners) and true negatives (i.e. non-corners), a pre-process based on straw graph is used to identify candidate corners from all points. As shown in Figure 6a, the horizontal axis of a straw graph is point index, while the vertical axis is the straw value (defined in [6]). The straw value of a line-point is close to 1, while is less than 1 for a curve-point. For a high-curvature point, it will deviate from 1. A corner is a point with local minimal straw. Thus, supposing a bug is walking along the straw curve. It first find an entry point where its straw value is a little smaller than 1. In our experiment, 0.995 is used which is very close to 1.0. Our purpose is to keep every real corners in our candidate corners. The critical problem is how to determine the exit point. The straw value of the exit point is computed by

$$straw_e = 0.4 \cdot straw_{min} + 0.6 \qquad (1)$$

Where $straw_{min}$ is the local minimal straw value between an entry and exit point; when $straw_{min}$ is updated by a new curve-point, the $straw_e$ is updated, too. The exit straw value is dynamic. When the current point's straw value is equal to greater than $straw_e$, the bug will exit in its next point. In such case, the point with local minimal straw value is selected as a candidate corner. The red and green points in Figure 6b are the candidate corners of this half-circle shaped stroke.

After this pre-process, users only need to label real corners out from these candidate corners by a simple click operation. This reduce the workload. If the type of a candidate corner is uncertain, such as the locations 1, 2, and 3 shown in Figure 4, five participants need to vote. If over 80% participants agree, the consensus is accepted. Otherwise, this stroke is removed.

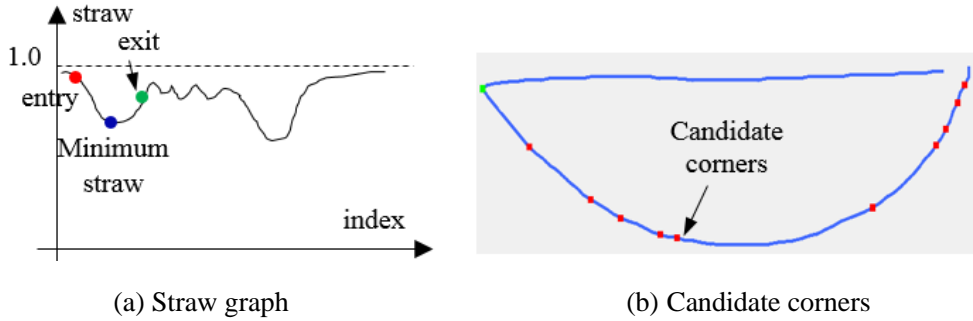

(a) Straw graph          (b) Candidate corners

Figure 6. Selects candidate corners based on straw graph

Third, line-points and curve-points are labelled. Other points except corners are labelled as line-points (green) and curve-points (blue). Each stroke has an orientation, i.e. the 128 points are ordered. To label a consecutive points, users first select a start and an end point by mouse click operation. Then, select the wanted point type in "Type" list box (Figure 5) and click "Add Label" bottom. We do emphasize that the points' types should be labelled from a larger point context, but we do not ask users to make all points between two consecutive corners the same type. That is, users can label points between two consecutive corners in different types.

In this way, 1500 strokes are labeled. The train set of multi-scaled context images for Corner-ResNet and Point-ResNet are generated according to these labels. In addition, the labels are used to evaluation the algorithm's performance in a cross-validation way.

**Multi-scaled context image generation**

ResNet is an image oriented residual neural network. It can learn from exemplar images. To be more accurate, we expect a point is judged from larger contexts. This is quite similar with human's perception. Meanwhile, we also hope its recognition ability can handle unknown strokes. This requires it must learn geometric properties from smaller contexts. Therefore, we generate four multi-scaled context images for a point.

Four scales 32*32, 64*64, 128*128, 256*256 are adopted, considering a balance between train accuracy and computational complexity. Before image sampling, input strokes are resized to 450*450 pixels. Figure 7 are four context images. All context images are resized to 224*224. Thus, the contour in 32*32 is a little blurred since it is enlarged, which is just a local area around the central point. The 256*256 context is almost one third of the whole stroke.

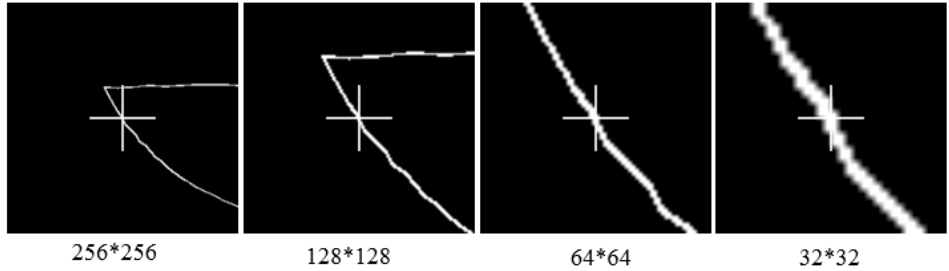

Figure 7. Multi-scaled context images

For Corner-ResNet, the central points are the candidate corners of a stroke, obtained by the pre-process step introduced in Figure 6. The number of candidate corners is about 3%~10% of the total points. Thus, the number of train images is equal to 900*4*|CadidateCorners|.

For Point-ResNet, the central points are all points of a stroke. Thus, the number of train images is equal to 900*4*128=460800.

**4.2. Corner and point classification**

This is the first step of the online phase. The input stroke is first resampled uniformly with 128 points. Then, the candidate corners are obtained by the pre-process step introduced in Figure 6. For each candidate corner, four multi-scaled context images are generated. These images are sent to Corner-ResNet to identify the true corners. The red points in Figure 8 are the true corners output by Corner-ResNet. However, the corner in Figure 8c is a false corner, which looks like a corner viewed from a small context image. Some of similar cases can be rectified in the segment merge process detailed in section 5.2.

Meanwhile, the multi-scaled context images of the input stroke's points are captured and sent to Point-ResNet. All non-corner points are classified into line-points or curve-points. As shown in Figure 8, line-points are displayed in green color, while curve-points are shown in blue color. Some of these points are classified in a wrong type, due to noise in a free hand-drawn stroke. Inspired by vote theory, similar voting scheme is adopted in our segment merge process. That is, if most points of a potential primitive agree with a primitive type, this type is accepted.

From these exemplar strokes, we know that a classified stroke usually contains many segments, separated by wrong classified points. Their lengths are different. A set of primitives should be identified from these segments to insert new corners and tangent points. Details are given in the following section.

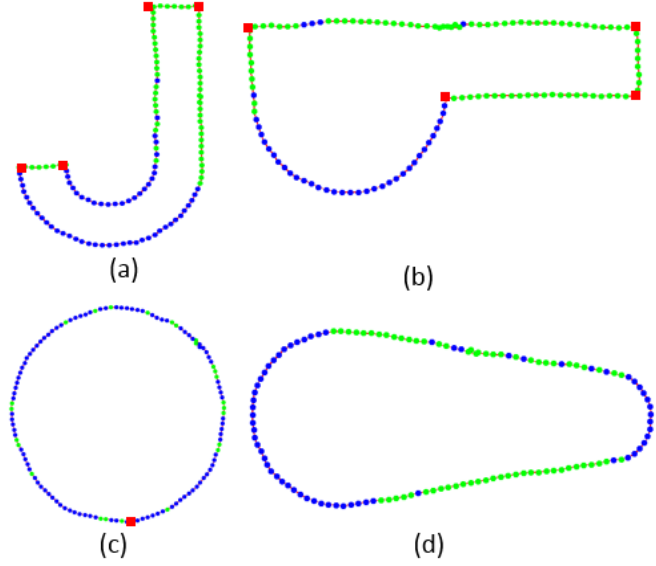

Figure 8. Corner and point classification results

## 5. Corner and tangent point detection

A stroke after classification with Corner-ResNet and Point-ResNet is composed of segments. The adjacency patterns of segments can be summarized, from which conditions and operations can be identified. These are the basis to design a robust segment merge algorithm.

### 5.1. Segment patterns

The input is a classified stroke, $stk = \{(x_i, y_i, l_i) | i = 1, \cdots, n\}$. That is, all points of a stroke are classified into line-points, curve-points, and corners with Point-ResNet and Corner-ResNet. A stroke is decomposed into primitives at feature points. Due to noise in a hand-drawn stroke and error in machine learning algorithm, not all labels $l_i$ are correct. Thus, a primitive may be decomposed into several segments. As the curved primitive shown in Figure 9 contains 20 segments. A *segment* is a set of sample points of the same type, denoted as *Seg*. |*Seg*| is the number of sample points in the segment *Seg*. A segment can contain only one classified point, e.g. $SS_{10}$.

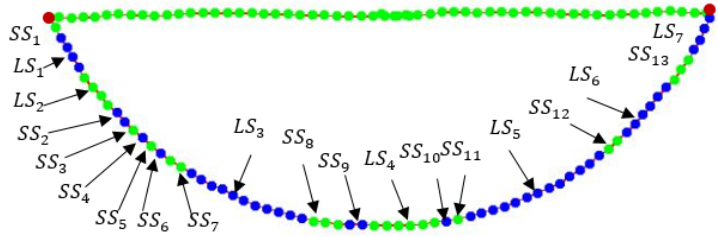

Figure 9. A curve primitive is decomposed into 20 segments due to point classification error (LS: long-segment; SS: short-segment; $\varepsilon_1 = 4$)

All segments are distinguished as long-segments and short segments. If $|Seg| < \varepsilon_1$, it is a short-segment, e.g. from $SS_1$ to $SS_{13}$ in Figure 9. Otherwise, it is a long-segment, e.g. $LS_1$ to $LS_7$. The threshold is a dynamic value which is proportional to the total number of points between two consecutive corners. The threshold should be less than 4 since we 4 consecutive points of the same type should be considered a long-segment. In our experiment, the proportional coefficient is 5%.

In a primitive, an internal segment has one left- and one right neighboring segment. A boundary-segment is a segment near a corner and it has only one neighboring segment. The distribution of short-segments and long-segments can be classified into four cases. Case 1 and case 2 are viewed from short segments, while case 3 and case 4 are considered from long-segments. Four cases are explained and their merge operation includes two steps: merge and update.

**Case 1**: a short-segment has only one neighboring long-segment.

For example, the short-segment $SS_{11}$ has a right long-segment $LS_5$. In this case, the short-segment can be merged directly into the long-segment. Then, the other neighboring segment of this short-segment is merged into this new merged long-segment. Taking the $SS_{11}$ for example, it will be merged by $LS_5$ and then the other neighboring segment $SS_9$ will be absorbed by the new long-segment ($SS_{10}$+ $LS_5$) in the update step.

The short-segments will be merged by different long-segment if we start from different short-segment. For example, if we start from $SS_{10}$, both short-segments $SS_{10}$ and $SS_{11}$ will be merged by long-segment $LS_4$. This problem will be resolved by the merge algorithm detailed in section 5.2.

**Case 2**: a short-segment has two neighboring long-segments.

For example, the short-segment $SS_{11}$ has two long-segments $LS_5$ and $LS_6$ neighbored. In such case, its two neighboring long-segments must belong to the same point type. According to the point type of the short-segment, two subcases exists.

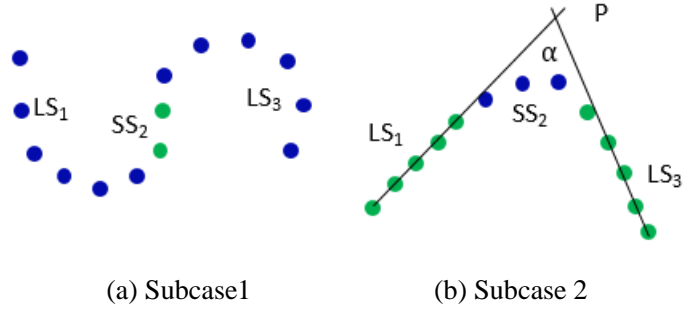

(a) Subcase1  (b) Subcase 2

Figure 10. Short-segment has two neighboring long-segments

**Subcase 1**: the short-segment is a line-point segment.

The short-segment $SS_{11}$ belongs to this subcase. Its neighboring long-segments must be curves. It usually occurs when users draw a curve. There exist two situations. If the curvature signs of the two long-segments are the same, e.g. $SS_{11}$ in Figure 9, the short-segment can be merged directly by the two long-segments. The new long-segment is the summation of points from the short-segment and two old long-segments.

If the curvature signs are opposite, e.g. $SS_2$ shown in S-shape stroke (Figure 10a), the points of the short-segment can be absorbed by a precision-lost test. First, for each long-segment, a least-square circle is fitted to five points near to the short-segment. Second, for each point of the short-segment, its distances to the two least-square circles are computed. Then, the point is merged into the long-segment with minimal distance. Finally, a tangent point is inserted between the two long-segments.

**Subcase 2**: the short-segments is a curve-point segment.

Its two neighboring long-segments must be straight lines. It usually happens when users draw a corner in a smooth way, as shown in Figure 10b. For each long-segment, a least-square fitting line is obtained first. Supposing their intersection point and angle are $p$ and $\alpha$, respectively. Then, if $\alpha$ is smaller than a threshold, the two long-segments are collinear, the short-segment is directly merged. Otherwise, a new corner will be inserted and the short-segment will be split. The split-point is the nearest point of the short-segment to point $p$. The split-point will be merged into which long-segments, depending its distances to two fitting lines.

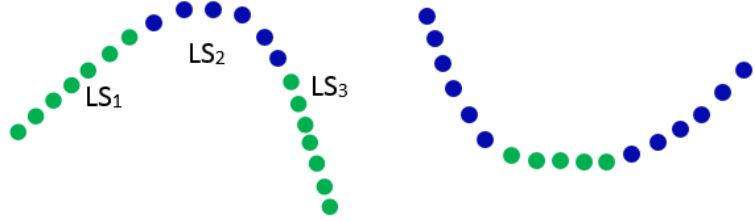

Figure 11. Long-segment has two neighboring longer long-segments

**Case 3**: a long-segment has two neighboring longer long-segments.

This often occurs when all short-segments have been merged. As shown in Figure 11, supposing the three consecutive long-segments are $LS_1$, $LS_2$, and $LS_3$, with $|LS_2|<|LS_1|<|LS_3|$. The two longer long-segments must be the same type. This case is similar with case 2 except the middle segment is a shorter long-segment, compared to short-segment in case 2. It also has two subcases according to the type of the shorter long-segments.

If $|LS_2|<0.25*|LS_3|$, this case is converted into subcases of case 2. That is, $LS_2$ will be merged by $LS_1$ and $LS_3$. Otherwise, the three long-segments are kept and two tangent points are added.

**Case 4**: a long-segment has only one longer long-segment.

This can be found when one long-segment is near a corner point. If the length of the shorter long-segment is less than 0.3 times the length of the longer long-segment, it will be merged by the longer long-segment. Otherwise, they are kept and one tangent point is inserted.

**5.2. Segment merge algorithm**

A segment merge algorithm should be designed to find all correct primitives, for all points shown in Figure 9 which belong to more than 10 segments. The merge algorithm is an iterative process based on four distribution cases discussed in section 5.1. Its input is a list of classified points between two consecutive corners. Its output is a list of corners and tangent points. This algorithm consists of four major steps.

Step1. Remove false corners by angle ratio.

A corner angle is formed by a pair of rays, whose end points slide along the stroke. Xiong et al. [7] observed that this angle will not change too much if it is a real corner. Thus, two far separated pair of rays will form two angles. If ratio of this two angles is less than a threshold, the corner output from Corner-ResNet is a real corner. Otherwise, it is a false corner.

Step2. Construct the adjacency information of all segments.

First, the dynamic threshold is computed and used to identify two segment sets: short- and long-segments. The long-segment set should be sort according to segment length in a descending way. Then, the adjacency of segment is created to guarantee that each segment has a predecessor and successor segment, except the boundary segments.

Step3. Merge short segments of case 1.

This is an iterative process, as shown in Figure 12. The input is the long-segment set (ab. L-Seg.). If the size of the long-segment set is zero, all points are set to the majority type of these points, and the merge algorithm is jump to step 4. Otherwise, starting from the longest long-segment, its neighboring segments (ab. N-Seg) are checked. If they belong to case 1, they are merged and updated. Then, if the new neighboring segments still belong to case 1, it continues to merge and update, until its two neighboring segments are long-segments. Then, the next long-segment starts to merge short-segments. In this way, a high priority is given to longer long-segments to merge short-segments. Until all long-segments in L-Seg are processed, step 3 is finished and go to step 4.

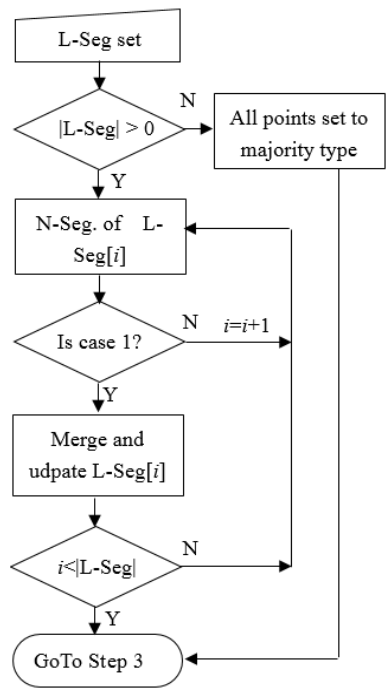

Figure 12. Flowchart of merge segments of case 1

Step4. Merge short segments of case 2.

During step3, the short-segments of case 1 have been processed. If there still short-segments left, it should belong to case 2. This step is a loop process. For each short-segment of case 2, the merge and update operation are performed according to two subcases of case 2.

Step5. Merge long-segments.

If there are more than one long-segments after step1-step4. They are sort in ascending order by their length. This step is also an iterative process. Starting from the shortest long-segment, it will be handled according to case 3 or case 4. This process is stopped when all long-segments have been processed.

**5.3. Segment merge example**

Two examples are given to show the work procedure of the segment merge algorithm. The first example is the noisy stroke shown in Figure 9. There are 69 points between the two consecutive corners (red points), which are decomposed into 20 segments. The threshold for short- and long-segments is equal to 4. The 20 segments are classified into 13 short-segments and 7 long-segments. The sorted long-segments are $LS_5$, $LS_3$, $LS_6$, $LS_4$, $LS_1$, $LS_7$, and $LS_2$. Here, $|LS_6|$ and $|LS_4|$ are equal, but $LS_6$ has the same type as longest long-segment $LS_5$, thus $LS_6$ is put before $LS_4$. The length of each segment is summarized in the first and second row of Table 1.

Table 1. Merge procedure for the classified stroke shown in Figure 9

| Seg | 1 | 2 | 3 | 4 | 5 | 6 | 7 | 8 | 9 | 10 | 11 | 12 | 13 | 14 | 15 | 16 | 17 | 18 | 19 | 20 |
|---|---|---|---|---|---|---|---|---|---|---|---|---|---|---|---|---|---|---|---|---|
| \|Seg\| | 1 | 4 | 4 | 2 | 1 | 1 | 1 | 1 | 2 | 11 | 3 | 2 | 6 | 1 | 1 | 13 | 2 | 6 | 3 | 4 |
| Step2 | 5 | 4 | | | | | 24 | | | | | | 6 | | 15 | | 2 | 6 | 3 | 4 |
| Step3 | 5 | 4 | | | | | 24 | | | | | | 6 | | | | 30 | | | |
| Step4 | 69 | | | | | | | | | | | | | | | | | | | |

The output of the step2, step3 and step 4 are listed in Table 1. In step 2, the yellow colored short-segments in 2$^{nd}$ row belong to case 1 and are merged into long-segments. That is, $SS_1$ is merged with $LS_1$; $SS_2$ to $SS_{11}$ are merged by $LS_3$; $SS_{12}$ and $SS_{12}$ are merged to $LS_5$. In step 3, the green colored short-segments in 3$^{rd}$ row belong to case 2 and are merged by $LS_5$, whose size grows to 30 points. In step 4, all five long-segments are merged into one single segment including 69 points.

Table 2. Iterative process of step2 for the classified stroke shown in Figure 9

| Line | L-seg | N-seg | Process |
|---|---|---|---|
| 1 | $LS_5$ | $SS_{11}$, $SS_{12}$ | $SS_{11}$ is case 1, $SS_{11}$ and $SS_{10}$ are merged; $SS_{12}$ is case 2. |
| 2 | $LS_5$ | $LS_4$, $SS_{12}$ | $LS_4$ is long-seg and $SS_{12}$ is case 2; $|LS_5|$=15. |
| 3 | $LS_3$ | $SS_7$, $SS_8$ | Both are case 1, $SS_7$, $SS_8$, $SS_6$, and $SS_9$ are merged. |
| 4 | $LS_3$ | $SS_5$, $LS_4$ | It is case, $SS_5$, $SS_4$ are merged. |
| 5 | $LS_3$ | $SS_3$, $LS_4$ | It is case, $SS_3$, $SS_2$ are merged. |
| 6 | $LS_3$ | $LS_2$, $LS_4$ | Both are long-segments; $|LS_3|$=24. |
| 7 | $LS_6$ | $SS_{12}$, $SS_{13}$ | Both are case 2; $|LS_6|$=6. |
| 8 | $LS_4$ | $LS_3$, $LS_5$ | Both are long-segments; $|LS_4|$=6. |
| 9 | $LS_1$ | $SS_1$, $LS_2$ | $SS_1$ is case 1, it is merged. |
| 10 | $LS_1$ | $LS_2$ | $LS_2$ is long-segment; $|LS_1|$=5. |
| 11 | $LS_7$ | $SS_{13}$ | $SS_{13}$ is case 2; $|LS_7|$=4. |
| 12 | $LS_2$ | $LS_1$, $LS_3$ | Both are long-segments; $|LS_2|$=4. |

The detailed iterative process of step 2 includes 10 steps, as shown in Table 2. It starts from the longest long-segment $LS_5$, which merges $SS_{10}$ and $SS_{11}$ (line1-2). Its length grows to 15 points. Then, $LS_3$ has the opportunity to merge short-segments. From line 3 to line 6, it merges shor-segments $SS_2$ to $SS_9$ and its length increases to 24 points. Also, the $SS_1$ is merged by $LS_1$. The long-segments $LS_6$, $LS_4$, $LS_7$, and $LS_2$ have nothing to merge.

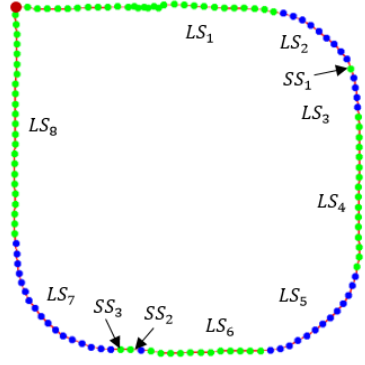

Figure 13. The 2nd example demonstrates the segment merge procedure

The second example is shown in Figure 13. The perfect stroke contains three arcs. Between the two adjacent corners (red points), the total number of points is 111. Thus, the threshold for short- and long-segment is 4 since 5%*111 is greater than 4. All points are decomposed into 8 long-segments and 3 short-segments. The total $SS_1$ is generated due to classification error, while $SS_2$ and $SS_3$ are mainly because of noisy drawing.

The merging procedure is listed in Table 3. In step 2, the yellow colored short-segments in 2nd row belong to case 1 and are merged into long-segments. That is, $SS_2$ and $SS_3$ are merged with $LS_7$. In step 3, the green colored short-segments in 3rd row belong to case 2. $LS_2$, $SS_1$ and $LS_3$ are merged into $LS_2$, whose size grows to 14 points. In step 4, all 7 long-segments are kept and 6 tangent points are inserted.

Table 3. Merge procedure for the classified stroke shown in Figure 13

| Seg | 1 | 2 | 3 | 4 | 5 | 6 | 7 | 8 | 9 | 10 | 11 |
|---|---|---|---|---|---|---|---|---|---|---|---|
| \|Seg\| | 14 | 9 | 1 | 4 | 16 | 13 | 12 | 1 | 2 | 17 | 22 |
| Step2 | 14 | 9 | 1 | 4 | 16 | 13 | 12 | 20 | | | 22 |
| Step3 | 14 | 14 | | | 16 | 13 | 12 | 20 | | | 22 |
| Step4 | 14 | 14 | | | 16 | 13 | 12 | 20 | | | 22 |

## 6. Experiments and discussion

In this section, the performance and generalization ability of our CTPD-DL are evaluated and compared to state-of-the-art CTPD algorithm. The stroke dataset has 1500 strokes of 20 shapes drawn by 15 users. It consists of two datasets, i.e. TCVD-1000 and CTPD-500. The former can be downloaded from TCVD-R and the latter is collected by our research team. The stroke database and labels can be downloaded for free. In our experiments, the dataset usually is grouped into train (60%), validate (20%), and test sets (20%). That is, the stroke dataset is divided. To consider the generalization ability of CTPD-DL to users and shapes, the stroke dataset are divided into 5 subsets by two strategies: set-by-persons and set-by-shapes (as shown in in Table 4). In set-by-person, the first subset contains 300 strokes drawn by 1st, 7th, and 15th persons. In set-by-shape, the first subset includes strokes of the 1st, 11th, 17th, and 19th template shapes.

ResNet is a residual neural network implemented with the deep learning framework PyTorch 0.4. The batch size is 128 and the number of epoch is 50. The training process is run on a GForce 1080Ti

GPU. Adam optimizer is used. The learning rate is $10^{-3}$.

Table 4. Five subsets divided by two different strategies

|  | Set-by-person | | | | |
|---|---|---|---|---|---|
| Index | 1 | 2 | 3 | 4 | 5 |
| Persons | 1, 7, 15 | 2, 8, 11 | 3, 9, 12 | 4, 10, 13 | 5, 6, 14 |
|  | Set-by-shape | | | | |
| Index | 1 | 2 | 3 | 4 | 5 |
| Shapes | 1, 11, 17, 19 | 2, 12, 18, 6 | 3, 13, 16, 9 | 4, 14, 15, 8 | 5, 20, 7, 10 |

In the following subsections, the effects of context images are explored in section 6.1. Then, the accuracy of Point-ResNet and Corner-ResNet are explored in section 6.2 and section 6.3, respectively. The performance of CTPD-DL is evaluated and compared in section 6.4.

### 6.1. Effect of context images

This section verifies the argument with experiment: points' type can be understood better if they are checked from multi-scaled context images. The Point-ResNet is select to test the scale effect of context images. The 1st subsets of both set-by-person and set-by-shape in Table 4 are selected as test sets; the 2nd, 3rd, and 4th subsets are the train set; and the 5th subset is the validate set. The image number is changed from one to four and the image scale is changed from 256*256 to 32*32, as the first column shown in Table 5. The accuracy (defined in section 6.2) is shown in Table 5.

Table 5. Effect of image size for Point-ResNet accuracy

| Train images | Set-by-persons | Set-by-shapes |
|---|---|---|
| 256*256 | 96.71% | 95.81% |
| 256*256, 128*128 | 96.62% | 95.66% |
| 256*256, 128*128 64*64 | 96.45% | 96.15% |
| 256*256, 128*128 64*64, 32*32 | 96.51% | 96.38% |

From this table, we know that the round with only one context image has the highest accuracy in the set-by-persons strategy. In set-by-persons, the Point-ResNet saw all shapes. Thus, it tends to remember those context images which were seen before in its training phase, to obtain a higher accuracy score. Thus, the round with only one context image has the highest accuracy since it is not affected by other scaled context images. However, for set-by-shapes strategy, some of the shapes are not seen before by Point-ResNet. Thus, beside global information, it still needs local data from smaller context images to distinguish a point from others. Thus, the round with four multi-scaled images has the highest accuracy because it also has the smallest context image, i.e. 32*32.

Thus, the experiments for Point-ResNet and Corner-ResNet adopt the four multi-scaled context images. When this tool is used in a SBEM system, its usage is similar with set-by-shapes, since most strokes are drawn in a free way by users, where the strokes are unknown shapes for the Point-ResNet.

### 6.2. Accuracy of Point-ResNet

The Point-ResNet classifies all points into line-points and curve-points. Its accuracy is computed by

$$Acc_p = \text{\#correctly classified points/\#total classified points} \quad (2)$$

The generalization ability of Point-ResNet should be investigated from both persons and shapes. A cross validation scheme is adopted. That is, this experiment is finished in 5 rounds, as shown in Table 6. In each round, a subset is selected as the test set; the neighboring subset is treated as the validate set; the other three subsets are the train set. For the 1$^{st}$ round experiment, i.e. the second column of Table 6, the 1$^{st}$ subset (defined in Table 4) is the test set; the 2$^{nd}$ subset is the validate set; the 3$^{rd}$, 4$^{th}$, and 5$^{th}$ subsets are the train set.

For a round of experiment, the number of context images of a point is 4 (optimized image scale is experiment in section 6.1). The total number of train context images is 460800. The training process of a round cost about 2.5 hours. The accuracy for 10 rounds for set-by-person and set-by-shape are listed in Table 6.

Table 6. Effects of persons and shapes on Point-ResNet's accuracy

| Test subset | 1 | 2 | 3 | 4 | 5 | Average |
|---|---|---|---|---|---|---|
| Set-by-person | 96.51% | 96.22% | 96.72% | 96.78% | 97.35% | 96.7% |
| Set-by-shape | 96.38% | 96.21% | 97.29 | 93.31% | 92.95% | 95.2% |

The set-by-persons experiment tests Point-ResNet's generalization ability for persons. The Point-ResNet sees all stroke shapes. The only difference between train and validate sets is those shapes are drawn by different users, i.e. different styles. From Table 6 we know that the accuracy is stable. This means draw styles have minor effects on the accuracy of Point-ResNet. Please note that, in each validate set, the proportion of strokes from TCVD-R subset and our subset is the same.

The set-by-shapes experiment explores Point-ResNet's generalization ability for shapes. The Point-ResNet only learns from part of the template shapes. From Table 6 we know the accuracy fluctuates a little wider. It has a lower accuracy when a test stroke contains some context images never seen by the Point-ResNet before.

### 6.3. Accuracy of Corner-ResNet

The Corner-ResNet identifies real corners from candidate corners. The output of Corner-ResNet is sent to segment merge algorithm (section 5.2), which will rectify two error cases. The first case is that a real corner is classified as a non-corner, i.e. false negative (FN). The second case is false corners, i.e. false positive (FP). Thus, the accuracy can be computed by:

$$Err_{FN} = 1 - R_c = \frac{FN_c}{TP_c + FN_c} \quad (3)$$

$$Err_{FP} = 1 - P_c = \frac{FP_c}{TP_c + FP_c} \quad (4)$$

Where *TP, FP,* and *FN* are true positives, false positives, and false negatives, respectively.

The Corner-ResNet has the same hidden layer structure with Point-ResNet. The only difference is the train context images. In this experiment, the first subset of the set-by-shape and set-by-person in Table 4 are selected. The number of candidate corners are 2633 and 3166 after the pre-process (introduced in section 4.1), for set-by-persons and set-by-shapes, respectively. When these candidate corners are sent to Corner-ResNet, the statistical data are listed in Table 7.

Table 7. Statistical data of Corner-ResNet when the 1st subset is selected as the test set

|  | #Candidate | TP | FP | TN | FN | $R_c$ | $P_c$ | $Err_{FN}$ | $Err_{FP}$ |
|---|---|---|---|---|---|---|---|---|---|
| Set-by-persons | 2633 | 882 | 5 | 1734 | 3 | 99.66% | 99.44% | 0.34% | 0.56% |
| Set-by-shapes | 3166 | 850 | 7 | 2258 | 7 | 99.18% | 99.18% | 0.82% | 0.82% |

From this table, we know the precision and recall accuracy for both strategies are over 99%. Similarly, the accuracy for set-by-persons is a little higher than set-by-shapes. However, the number of false negatives and false positives are critical for the accuracy of CTPD-DL. Take set-by-shapes strategy for example, 7 false positives (FP) appears in 4 strokes of the 17th stroke (Figure 14a) and in 3 strokes of the 19th stroke (Figure 14b), respectively; all 7 false negatives (FN) are found in 7 strokes of one shape, i.e. the 1st shape (Figure 14c). These two wrong types of corners will be rectified in the segment merge process.

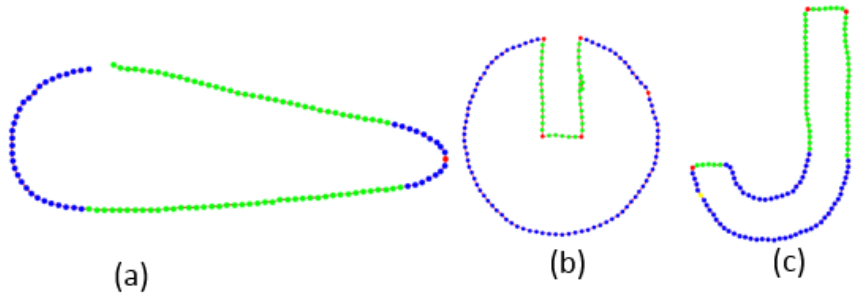

Figure 14. Exemplar strokes with FN and FP points

### 6.4. Performance of CTPD-DL

In this section, the accuracy, generalization ability and time cost of CTPD-DL algorithm are evaluated and compared to the state-of-the-art algorithm TCVD-R. The all-or-nothing (AON) accuracy defined by Wolin et al. [6] is adopted.

$$AON = \#correctly\ detected\ strokes / \#total\ strokes \qquad (5)$$

**Accuracy comparison**

The accuracy of the CTPD-DL and TCVD-R is compared on two datasets. First, the dataset TCVD-1000 is adopted since the TCVD-R is optimized on this dataset. The AON accuracy of TCVD-R is obtained by testing the 1000 strokes with the TCVD-R executable program. To be as fair as possible, 600, 200, and 200 strokes are selected by person as the train, validate, and test sets for our algorithm, respectively. The AON accuracy is shown in the 2nd column of Table 8. Please note that the accuracy 91.5% for TCVD-R is smaller than their published result 94.3%. This is because we compare the detected corners and tangent points to users' labels, not the template shapes. From this table, the AON accuracy of CTPD-DL is higher than that of TCVD-R.

Table 8. Comparison to state-of-the-art algorithm

| Approach | TCVD-1000 | TCVD-1000+ CTPD-500 |
|---|---|---|

|  |  | Set-by-person | Set-by-shape |
|---|---|---|---|
| CTPD-DL | 97% | 97.0% | 95.0% |
| TCVD-R | 91.5% | 84.6% | |

Second, five round experiments (the same with that of the Point-ResNet listed in Table 6) are performed with a cross validation scheme. The averaged AON accuracies are listed in the 3rd and 4th columns in Table 8. The AON accuracy of TCVD-R is obtained by testing the 1500 strokes with the TCVD-R executable program.

From this table, we observed that the AON accuracy is a little higher than $Acc_p$ listed in Table 6. This is because the false positive points (i.e. false corners) are removed by step 2 and the lost corners (i.e. FN) are added in step 3 and step 4 of segment merge algorithm (detailed in section 5.2). However, there are still 14 wrong detected strokes. The cause for these errors are analyzed and summarized into four reasons, as listed in Table 9.

Table 9. Four reasons for failed strokes of CTPD-DL

| No. | Reasons | Strokes | Number |
|---|---|---|---|
| 1 | False positive point | 3-1-19, 3-3-19, 10-5-17, 11-5-17, 15-5-17, 4-5-19, 7-4-17 | 7 |
| 2 | False negative | 3-1-1, 3-5-1, 4-2-1, 7-3-1, 15-4-1 | 5 |
| 3 | Threshold value | 4-2-19, 4-5-19 | 2 |
| 4 | Stroke end is overlap | 2-3-11 | 1 |
|  | **Total** |  | **15** |

The representative error strokes are shown in Figure 15. The 1st reason is the false positive point. All these errors can be found in the 17th and 19th template shapes. Two exemplar strokes are given in Figure 15a and Figure 15b, respectively. This is because the false positives generated by Corner-ResNet cannot be removed by step 2 of segment merge algorithm. The parameter value of iStraw are empirical-based. It is hard to tune well for all strokes. The 2nd reason is false negative, i.e. real corners are lost in the Corner-ResNet. We observed the 5 wrong strokes are all of the 1st template shape. Figure 15c is one of the representative stroke. The 3rd reason is because of the threshold value used in step 4 of the segment merge algorithm. In Figure 15d, the left 4 curve-points is greater than 25% of the line-points, thus it is kept as a long-segment, where a tangent point is added to their transition point. Thus, this is a wrong stroke. The 4th reason is because the stroke end has overlap area. Since the Point-ResNet does not know these points are actually separated, it still judges a point from their context images. Thus, most of them are classified as curve-points, leading to this wrong stroke.

We compare our algorithm with the state-of-the-art TCVD-R algorithm. It is selected for three reasons. First, it can detect both corners and tangent points, too. Second, it published their 1000 strokes for free download. Third, they also published an executable program for research test. Thanks them for their generosity.

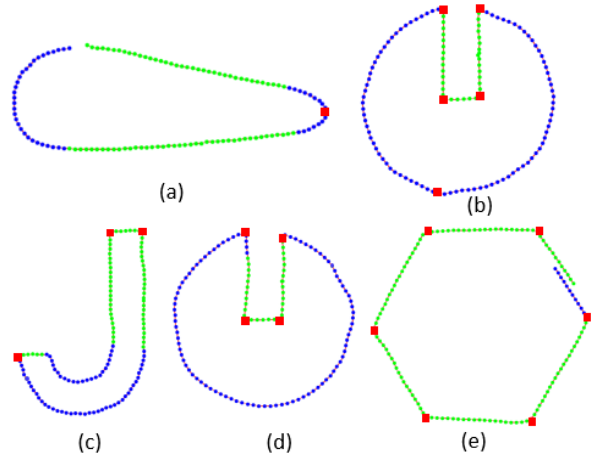

(a)  (b)  (c)  (d)  (e)

Figure 15. Error strokes of our algorithm

The TCVD-R is still a heuristic-based approach. The parameters of the published TCVD-R executable program are optimized with 200 strokes, which is formed with a stroke of each shape for each participants. That means their optimization process saw every template shapes and all drawing styles. Thus, we plan to test their AON accuracy on the 1500 strokes, though 200 of them are used to train their parameters. The major difference compared to their original experiment is we evaluate the detection results with users' labels, not the template shapes.

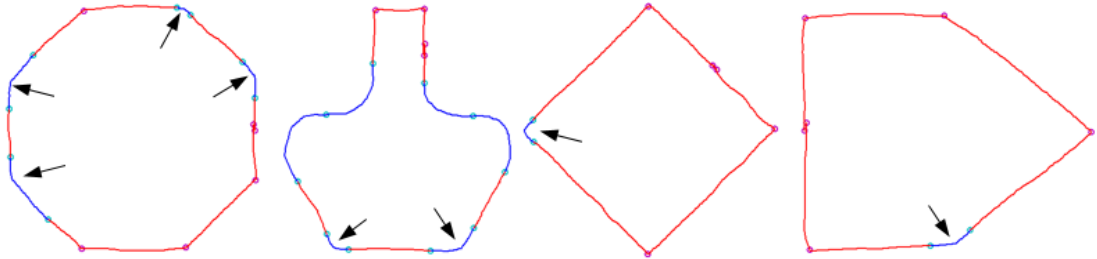

Figure 16. Error strokes of TCVD-R algorithm

Its AON accuracy is 84.6%. The representative error strokes are shown in Figure 16. As you can see, the major error type is that TCVD-R treats little smooth corners into curve segments. Thus, a corner is replaced with two tangent points. This interrupts users' design process since they need to stop current drawing process and rectify these errors. This will deteriorate user experience.

**Generalization ability comparison**

Besides to the generalization ability to persons and shapes, this section will check the generalization ability to datasets of different distributions, i.e. TCVD-1000 and CTPD-500.

To compare the generalization ability, the CTPD-500 subset is selected as the test set. For TCVD-R, we convert the CTPD-500 strokes into TCVD's format and resize to their scale. They are imported to TCVD-R executable program and the detection results are checked manually. If the detected corners and tangent points are the same with the stroke's labels, it is right; otherwise, it is wrong. The final AON accuracy is 71.4%.

For CTPD-DL, 800 strokes of TCVD-1000 is selected as the train set with set-by-person strategy. Similarly, 200 strokes of TCVD-1000 are left as the validate set. The test set is the CTPD-500 strokes. The accuracy of CTPD-DL is 88.6%, which is much higher than TCVD-R.

**Time cost**

In our experiment, the time cost is 0.37s per shape using CTPD-DL with four multi-scaled context images. This time cost includes three parts: corner detection with Corner-ResNet (14.2%), point classification with Point-ResNet (84.7%), and segment merge algorithm (1.1%). The Point-ResNet is the most time-consuming part because the Point-ResNet analyzes 512 context images during its classification process. Corner-ResNet and Point-ResNet run on a GForce 1080Ti GPU and the segment merge algorithm run on an Intel i7 CPU PC with Windows 10 operating system.

## 7. Conclusion and future work

This paper proposed a corner and tangent point detection algorithm for hand-drawn strokes with residual neural network—ResNet. Its robustness to different persons, stroke shapes, and biased dataset was improved from two aspects. First, multi-scaled point context images were adopted to classify a stroke's points. It is more accurate if a point is viewed from a larger point context. Second, a vote scheme was used to determine the type of a primitive. The segment patterns, conditions, and merge operations were analyzed, which were used to design a robust segment merge operation. This is tolerant to wrong classification of points. Our approach consists of two steps. First, each point of a stroke is classified as corners, or line-points, or curve-points by two residual neural networks, i.e. Corner-ResNet and Point-ResNet. In such way, a stroke is decomposed into adjacent segments. Then, the adjacent short- and long-segments are merged into primitives, where new corners and tangent points are inserted at transition points between primitives.

In the experiments, 1500 strokes drawn by 15 participant in 5 rounds are separated into train set, validate set and test set. A cross validate scheme is adopted for various experiments. We found that four multi-scaled context images has the best accuracy for practical usage. Under set-by-person and set-by-shape strategies, the AON accuracies of CTPD-DL are 97% and 95%, respectively, compared to 84.6% of the state-of-the-art TCVD-R technique. The datasets with different distribution, the AON accuracy of CTPD-DL is 88.6%, compared to 71.4% of TCVD-R. That is, the CTPD-DL technique has a better performance if it is used in different scenes.

In the near future, three aspects need to be improved to perfect CTPD-DL. First, more complex strokes and more participants should be used to test its accuracy and robustness. Second, an optimization method should be design to save the time cost of Point-ResNet. Third, some error types can be resolved or improved in the segment merge algorithm in the near future.

## Acknowledgement

This work was supported by the National Natural Science Foundation of China under Grant nos. 61502263.This work was supported by the National Natural Science Foundation of China under Grant nos. 61502263.